# Elasticity Maps of Living Neurons Measured by Combined Fluorescence and Atomic Force Microscopy

Elise Spedden*†, James D. White*†‡, Elena N. Naumova §, David L. Kaplan‡, Cristian Staii*†

* Department of Physics and Astronomy, Tufts University, †Center for Nanoscopic Physics, Tufts University, 4 Colby St, Medford, MA, USA; ‡Department of Biomedical Engineering, and Department of Chemical Engineering, Tufts University 4 Colby St, Medford, MA, USA, § Department of Civil and Environmental Engineering, Tufts University, 200 College Avenue, Medford, MA, USA

**ABSTRACT**

Detailed knowledge of mechanical parameters such as cell elasticity, stiffness of the growth substrate, or traction stresses generated during axonal extensions is essential for understanding the mechanisms that control neuronal growth. Here we combine Atomic Force Microscopy based force spectroscopy with Fluorescence Microscopy to produce systematic, high-resolution elasticity maps for three different types of live neuronal cells: cortical (embryonic rat), embryonic chick dorsal root ganglion, and P-19 (mouse embryonic carcinoma stem cells) neurons. We measure how the stiffness of neurons changes both during neurite outgrowth and upon disruption of microtubules of the cell. We find reversible local stiffening of the cell during growth, and show that the increase in local elastic modulus is primarily due to the formation of microtubules. We also report that cortical and P-19 neurons have similar elasticity maps, with elastic moduli in the range 0.1-2 kPa, with typical average values of 0.4 kPa (P-19) and 0.2 kPa (cortical). In contrast, DRG neurons are stiffer than P-19 and cortical cells, yielding elastic moduli in the range 0.1-8 kPa, with typical average values of 0.9 kPa. Finally, we report no measurable influence of substrate protein coating on cell body elasticity for the three types of neurons.

**INTRODUCTION**

In the developing brain neuronal cells extend neurites (axons and dendrites), which navigate and make connections with other neurons in order to wire the nervous system. The outgrowth of neurites from the cell body of a neuron is a highly complex process involving interactions with an inhomogeneous and changing extracellular environment (1, 2), detection and interpretation of multiple biochemical and geometrical cues (1-6), activation of many different transduction pathways (1, 2, 7, 8), and several types of intracellular polymerization-depolymerization processes (1, 7-10). Mechanical interactions and physical stimuli play a key role in many of these processes whether one considers the rearrangements of the cytoskeleton and the generation of traction forces as a result of neurite growth, the adhesion of neurites to extracellular matrix proteins, the change in orientation and velocity of the growth cone in response to guidance cues, or the axonal navigation over tissues of varying stiffness (11-15).

Knowledge of various mechanical parameters such as the elastic properties of the cells and the growth substrate, or adhesion forces and traction stresses generated during axonal extensions are therefore essential for a deep understanding of the mechanisms that control neuronal growth and development. For example, recent studies have also shown that substrate stiffness plays an important role in the growth of peripheral dorsal root ganglion (DRG) neurons (16). During neurite outgrowth DRG cells generate relatively large adhesion forces and traction stresses, and they also display a large degree of sensitivity to substrate stiffness, showing maximal outgrowth on substrates with elastic modulus of the order of 1 kPa. It was hypothesized that these strong neurite-substrate mechanical couplings enable DRG neurons to grow very long axons and also to sustain relatively large external forces exerted by the surrounding tissue (16). Other groups have reported that glial cells display maximum growth on even stiffer substrates of the order of several kPa (17-19). In contrast to the mechanical response displayed by DRG neurons and glial cells, primary cortical and spinal cord neurons have been reported to grow well on softer substrates with elastic moduli on the order of a few hundred Pa, comparable to the average stiffness of central nervous system (CNS) tissue (16, 18, 20). Moreover, several studies have shown that in general, CNS neurons are much less sensitive to substrate stiffness than peripheral neurons or glial cells (16, 21). It was argued that this difference in mechanosensitivity between glial cells, cortical neurons, and DRG neurons could play an essential role in the initial structuring of the nervous system (15).

When studying neuronal cells and other constituents of the nervous tissue (glial cells, extracellular matrix proteins etc.) one has to take into account that these are heterogeneous, viscoelastic materials and that their mechanical response depends on the timescale, magnitude and loading rates of the externally applied forces (13, 19, 22). Many experimental techniques have been used to measure mechanical responses from cells and growth substrates, including traction force microscopy (16, 23), optical and magnetic tweezers (24, 25), microneedle pulling (13, 26), coated micro beads pulling (27, 28) and Atomic Force Microscope (AFM) based nanoindentation (29-34). The particular capabilities of the AFM, such as nanometer-scale spatial resolution and positioning on the cell surface, high degree of control over the magnitude (sub- nN resolution) and orientation of the applied forces, minimal sample damage, and the ability to image and interact with cells in physiologically relevant conditions make this technique particularly suitable for measuring mechanical properties of living neurons.

Previous studies using AFM or other methods suggest that the mechanisms of neurite outgrowth and cytoskeletal dynamics in response to mechanical properties of the surrounding environment are extremely complex and that different types of neurons respond distinctly to the same physical cues. Furthermore, there is growing evidence that the elastic properties of cells are important for mechanosensitivity and that they are strongly correlated with cellular health, development and function (15, 29, 33, 35, 36). For example, recent AFM measurements have revealed significant quantitative differences between the mechanical properties of cancerous and healthy cells (36). AFM force spectroscopy combined with bulk rheology measurements have shown that CNS glial cells are softer than the surrounding neuronal tissue, suggesting that, at least in certain areas of the CNS glial cells act as a soft shock absorbing tissue, which protects neuronal cells in the case of mechanical trauma (22). Moreover, it was also reported that radial glial cells, along which neuronal cells grow during the initial stages of development, have mechanical properties that vary significantly between different regions of the CNS (15, 22, 37). Despite the fundamental role played by the interactions between mechanical stimuli and cell elastic properties during neuronal growth and development, currently there are no systematic studies that show how the intrinsic mechanical properties of the neurons change during growth, how the cell elasticity and stiffness vary between different types of neuronal cells, or how the variations in cellular elastic properties are related to differences in the local environment faced by different types of neurons.

To gain new insight into neuronal cell mechanics and outgrowth, the goal of the present study was to use combined AFM imaging, AFM force spectroscopy and Fluorescence Microscopy to produce systematic, high-resolution elasticity maps for three different types of live neuronal cells: cortical neurons obtained from rat embryos, chick dorsal root ganglion (DRG), and neurons derived from P-19 mouse embryonic carcinoma stem cells. These types of neurons are representative for: a) cells that grow in CNS (cortical), which typically interact with soft environments; b) the peripheral nervous system (DRGs), which face stiffer environments; and c) stem cell derived neurons (P19), which are measured for comparison with the other two cell types. We also measure how the elastic properties of each type of neuronal cell are influenced by the cell interaction with three different growth factors: poly-D-lysine, laminin and fibronectin. Furthermore, by taking advantage of the ability of AFM to both image and apply controllable forces to live cells over time we monitor how the dynamics of axonal growth affect the stiffness maps of neuronal cell bodies, and how the cell stiffness changes upon chemical modification (disruption of microtubules) of the cell. We present the first use of AFM elasticity mapping to monitor differences in neuronal cell body elasticity over time, resolving internal changes to live and healthy cells due to neurite extension and drug response. We also find support for DRG neurons interacting with their surrounding environment via larger forces and stresses than cortical cells.

## MATERIALS AND METHODS

**Surface preparation, cell culture and plating**

Cells were cultured on 3.5 cm glass disks designed to fit in the Asylum Research Bioheater fluid cell (Asylum Research, Santa Barbara, CA). Poly-D-lysine (PDL) (Sigma-Aldrich, St. Louis, MO) coated plates were made by immersing the disks in a PDL solution (0.1 mg/ml) for 2 hours at room temperature. The plates were rinsed twice with sterile water, and sterilized using ultraviolet light for ≥30 minutes. AFM disks were similarly coated with laminin (LN) or fibronectin (FN). LN plates were coated with 50 μg/ml natural mouse laminin (Life Technologies, Grand Island, NY) solution in buffered saline for 1 hour at 37°C. FN coated plates used a 0.1 mg/ml bovine plasma fibronectin (Life Technologies) solution in buffered saline for 2 hours at 37°C.

Dorsal root ganglia (surgically isolated from day 9 chick embryos), rat cortical neurons (obtained from embryonic day 18 rats), and P-19 neurons (obtained from mouse teratocarcinoma stem cells) were plated and incubated following standard procedures (See Text S1 in the Supporting Material for details). For all cell types, immunostaining experiments have indicated cultures of high neuron purity (see See Figure S1 in the Supporting Material for details).

**Force map acquisition and data analysis**

Force Maps were taken using an Asylum Research MFP3D AFM (Asylum Research, Santa Barbara, CA) with an inverted Nikon Eclipse Ti optical microscope (Micro Video Instruments, Avon, MA). The samples were mounted in an Asylum Research Bioheater chamber with cell culture medium and maintained at $37^0$ C during all experiments. All measurements were taken using Olympus Biolever cantilevers (Asylum Research, Santa Barbara, CA) with nominal spring constant of .03 N/m. Before measurement on a new sample, each cantilever was calibrated both in air and in the sample medium. 16 X 16 μm maps of individual force vs. indentation curves were taken on each cell with a resolution of 1μm between points (See Text S2 in the Supporting Material for details).

The elastic modulus values were determined by fitting the Hertz model for a 30 degree conical indenter to the acquired force vs. indentation curves using the Asylum Research MFP-3D Hertz analysis tools (See Text S2 in the Supporting Material for additional details). These values can be combined with surface height information to produce a topographical rendering with elastic modulus values mapped on the surface (see Fig. 1 *a*).

For the cell type and surface coating data (see Fig. 1 and Fig. 2), each map was characterized by 3 values. The top 10% of elastic modulus values measured on an individual cell were averaged to obtain the "Top 10%" (henceforth referred to as the "*highest values*") for that cell. The middle 30% of values from that same cell were averaged to obtain the value for "Middle 30%" for that cell (henceforth referred to as "*medium values*"). Lastly, the lowest 10% of values were averaged to obtain the "Bottom 10%" (referred to as "*lowest values*"). These three values allow a simple way to compare stiffness between maps of different cells (see Figure S2 in the Supporting Material for additional details). A typical force point has a fitting error of ≤ 20% and is typically repeatable to within this error. For the cell type comparison graph (see Fig. 1 *d*), the highest, medium, and lowest values for all cells of each type were averaged. The error bars represent the standard deviation of the mean. For the cell dynamics and drug study data, each map was characterized by a histogram of all points measured over the cell body. Each histogram was binned in groups of 200 Pa per bin. As each map varies slightly in the total number of points taken above the cell, these histograms were plotted as the percent of total measured map points per bin rather than total number of points. In this way the stiffness distributions of two different maps of the same cell with a slightly different total number of points may be directly compared.

All measured cells from all three types (cortical, P19, and DRG) had similar soma size, with an average diameter of (13±4) μm.

**Fluorescence microscopy**

Two types of fluorescent dyes were used in this study. For microtubule staining, the live cortical samples were rinsed with phosphate buffered saline (PBS) and then incubated at 37ºC with 50 nM Tubulin Tracker Green (Oregon Green 488 Taxol, bis-Acetate) (Life Technologies, Grand Island, NY) in PBS. The samples were then rinsed twice and transferred to the AFM Bioheater chamber and maintained at 37ºC while bright field and fluorescence images (using a standard Fluorescein isothiocyanate -FITC filter: excitation/emission of 495 nm/521 nm) were taken before and after each AFM elasticity map. All cells were measured within the first 2 hours on the AFM stage.

For F-actin staining, each set of live cortical cells was optically located on the pre-marked surface, and force maps were acquired for a small number (2- 3) cells. Immediately after the last map was taken, the sample was removed from the AFM stage and fixed in 10% Formalin for 15 minutes. The sample was then rinsed twice with PBS and permeabilized for 10 minutes with 0.1% Triton-100 (Sigma-Aldrich, St. Louis, MO) in phosphate-buffered saline (PBS) (Life Technologies, Grand Island, NY). The sample was then incubated at room temperature for 20 minutes in 50 μM Alexa Fluor® 564 Phalloidin (Life Technologies, Grand Island, NY), and rinsed with PBS. The same cells mapped via AFM were re-located optically using the sample marking and imaged fluorescently using a standard Texas Red (excitation/emission of 596/615) filter.

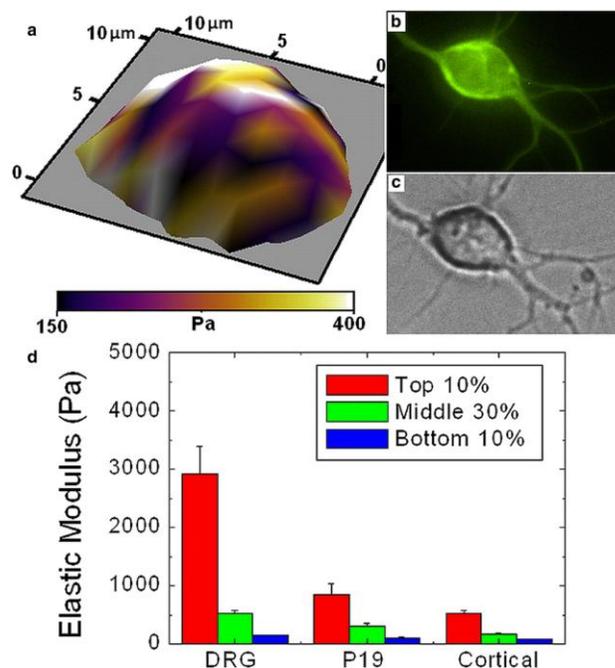

FIGURE 1 (a) 3-D topographical rendering of live cortical neuron body with color/shade indicating elastic modulus (Pa). (b) Optical fluorescence image of cell in shown in (a) stained for microtubules with 10nM Tubulin Tracker Green. (c) Bright field optical image of cell shown in (a, b). (d) Averages over all highest values (top 10%), medium values (middle 30%), and lowest values (bottom 10%) obtained from individual force maps of chick DRG, mouse P-19, and rat cortical neurons. Error bars represent standard deviation of the mean. Cortical and P-19 somas present significantly different highest, medium, and lowest values from DRG's (p≤0.001 one way ANOVA). All cells were measured in the passive (no neurite extension) state (see text).

## RESULTS

### Comparison between DRG, P-19 derived, and cortical neurons

Force maps were performed on cultures of each neuronal cell type, with a minimum of 15 cells examined in each data set to provide statistical significance for our results. For each cell we define an average highest, medium and lowest value for the elastic modulus, which allow us a simple way to compare stiffness between maps of different cells (see Force Map acquisition and data analysis section, and Figure S2 in the Supporting Material). Fig. 1 *d* shows that the average highest, medium, and lowest elasticity values for the P-19 derived neurons are similar to those obtained on cortical neurons despite the large morphological variation present in the P-19 derived neuronal cells. P-19 neurons (15 individual cells) yield an average highest value for the elastic modulus of (854 ± 181) Pa, compared with (521 ± 52) Pa for cortical neurons (24 individual cells). Similarly, the respective average value for the medium stiffness region is (301 ± 61) Pa for P-19 and (163 ± 15) Pa for cortical, and the average values for the lowest stiffness region are (104 ± 15) Pa (P-19) and (82 ± 5) Pa (cortical). All uncertainties in the values quoted here and in the following sections are standard deviations of the mean. All measured highest, medium, and lowest values of elastic moduli for cortical neurons fall in the range of the corresponding values measured on P-19's (with a significantly larger standard deviation of the mean present in the P19 sample set). These results support previous studies that indicate P-19 derived neurons exhibit some measureable characteristics of cortical region cells (32, 38).

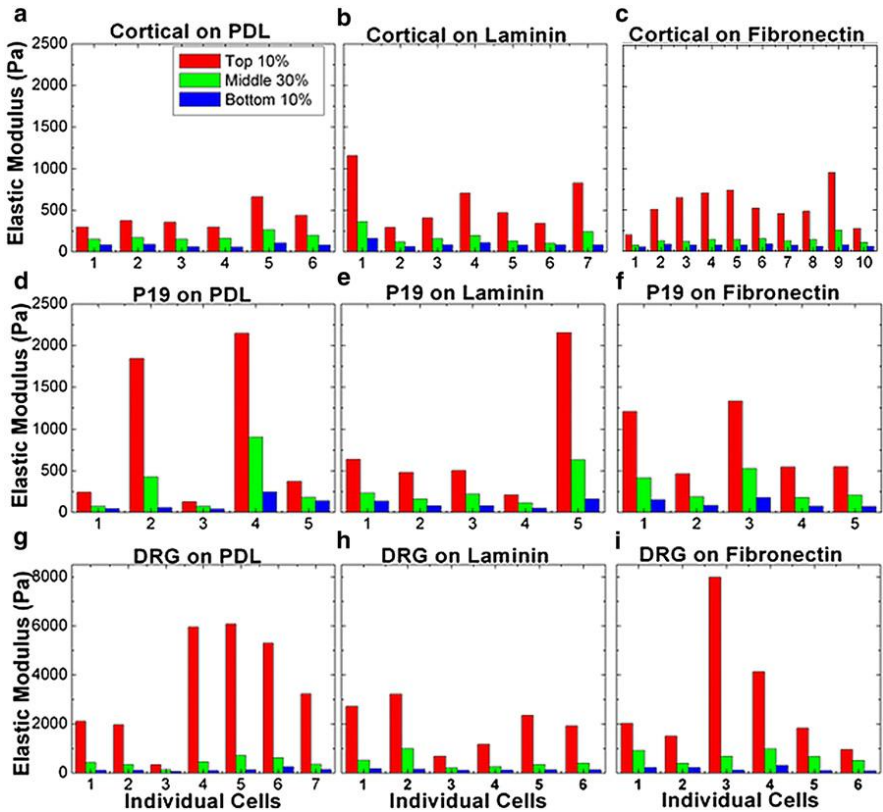

FIGURE 2 Elastic modulus values measured by AFM for individual cells (average of top 10% of values for each cell, average of middle 30% of values for each cell, and average of lowest 10% of values for each cell). The x-axis denotes individual cells. (a-c) Cortical neurons on (a) Poly-D-Lysine (PDL), (b) laminin (LN), and (c) fibronectin (FN). (d-f). P-19 neurons on PDL, LN, or FN, respectively. (g-i) DRG neurons on PDL, LN, or FN, respectively. All cells were measured in the passive (no neurite extension) state (see text).

DRG neurons yield significantly stiffer highest and medium elasticity values than P-19 or cortical neurons (Fig. 1 *d*). DRG's (19 individual cells) yield an average highest stiffness of (2920 ± 480) Pa, an average value for the medium stiffness region of (524 ± 58) Pa, and an average lowest stiffness of (144 ± 15) Pa (typical average value for the entire cell being ~ 900 Pa). These values are significantly different than the respective measured values for P-19 and cortical neurons (p≤0.001, one-way ANOVA). The average values yielded by the DRG neurons also fall around the range of the optimal substrate stiffness for DRG growth (1000 Pa) (16) indicating a possible stiffness match between cell body and optimal growth environment. The lowest stiffness values of the DRGs are very close to the corresponding lowest values of the generally much softer P-19 and cortical neurons. The regions of the cells corresponding to these values likely correlate to some similar soft internal components, such as fluid components of the cytoplasm.

**Effect of surface coating on cell body elasticity**

Many reports have shown that surfaces coated with different types of growth factors or extracellular matrix proteins can produce very different growth and adhesion dynamics for neuronal cells (1, 28, 39). Here we explore the effects of surface coating on the biomechanical properties of neuronal cells. Specifically, we take 1 µm resolution elasticity maps of P-19, cortical and DRG neurons plated on glass disks, each coated respectively with laminin, fibronectin, and poly-D-lysine (PDL). These are the most common types of proteins used in literature for *in vitro* neuronal culture. Fig. 2 shows the elastic modulus data collected for all the three cell types and surface coatings. All measured neuron types: cortical (Fig. 2 *a-c*), P-19 (Fig. 2, *d-f*) and DRG (Fig. 2 *g-i*) display significant variations in stiffness values among individual cells. For each cell type these cell-to-cell variations in stiffness are larger than the measured variations due to surface coating. 1-way ANOVA tests (see Table S1 in the Supporting Material) indicates no measureable correlation between variation in elasticity values and surface coating for the majority of combinations of neuronal cell type and surface coatings. In addition, for each cell type we have calculated cumulative distributions of measured elastic moduli for individual cells (see Fig. S2 *b* in the Supporting Material), as well as average cumulative distributions for each surface (Fig S2 *c, d*). This data shows that the variations among average cumulative distributions for each surface are *smaller* than the calculated standard deviations (see Fig S2 in the Supporting Material), further confirming a low probability for an effect due to surface coating outside one standard deviation.

**Cytoskeletal dynamics measured by combined AFM and fluorescence microscopy**

In an effort to better understand the regions of high stiffness on cortical neurons we used combined AFM and fluorescence microscopy to monitor the dynamics of these regions and characterize their underlying components. We have chosen cortical neurons to perform dynamics and fluorescence studies since this cell type has reproducibly shown both active and passive growth states (defined in the next section). To verify map consistency a number of N=6 cells that underwent no neurite growth were mapped multiple times (between 20 minutes to 2 hours apart). Average elastic modulus values for a given cell yielded consistent values, with cell-wide elastic modulus averages agreeing within 86% (see Fig. S3 in the Supporting Material).

**Effects of neurite length extension on cell body elasticity for cortical neurons**

Living cortical neurons on PDL coated glass disks were mounted in the AFM's temperature controlled Bioheater chamber on the second day after plating. The neurite extensions from these cells were observed in one of two primary states: an *active state*, where the growth cone was changing location and the neurite increased significantly in length ($> 5\mu m$) (compare Fig. 3 *a* and 3 *b*), or a *passive state*, where the growth cone was not observably active and the neurite length remained constant (compare Fig. 3 *c* and *d*). Neurites were seen to transition, over the course of 30 minutes to an hour, between passive and active states. In order to monitor the changes in cell body elasticity due to active neurite extension, elasticity maps were taken in succession (approximately 15 minutes for each map) on a given cell both before and during (or during and after) a phase of active neurite growth. Neurite extension was determined from the optical images of the cell taken before (Fig. 3 *a,c*) and after (Fig. 3 *b,d*) the acquisition of the force maps.

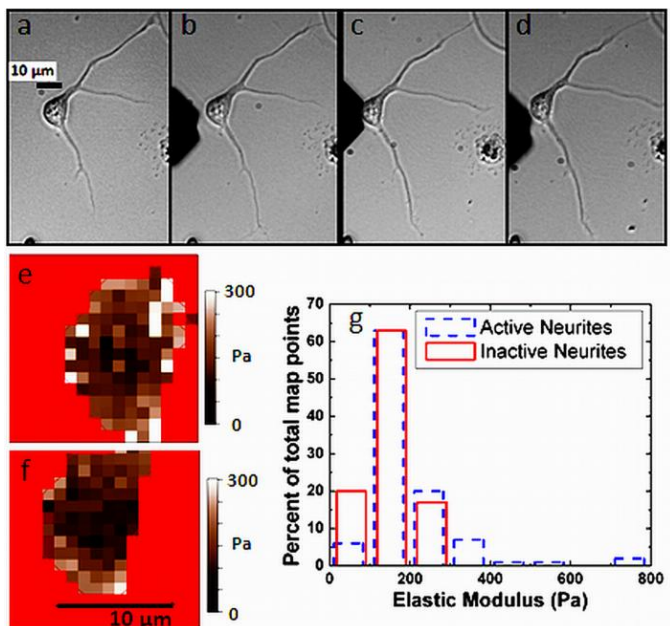

FIGURE 3 (a, b) Optical images before (a) and after (b) force measurements of a live cortical neuron undergoing active neurite extension during 15 minute force-map acquisition (i.e. change in growth cone position and morphology for the top neurite, increased in length for the bottom neurite). (c, d) Optical images before (c) and after (d) force measurements of same live cortical neuron at a later time *not undergoing* active neurite extension (passive phase). Scale bar shown in (a) is the same for all images (a-d). (e,) Elasticity map for the active extension phase shown in (a-b). (f) Elasticity map for the passive phase shown in (c-d). Scale bar shown in (e) is the same for both maps. (g) Histogram of percent of total map points in each elastic modulus bin (see Materials and Methods). Dashed line: data for active extension state. Solid line: data for the passive state. The average elastic modulus value increases by 35% during growth. Similar results were obtained on 4 additional cells.

During the active neurite extension phase we have measured an overall increase in the average values of the elastic modulus across the entire soma between 23% and 175%. In all cases (N=5 cells) the largest increase in stiffness (accounting for more than 75% of the observed overall increase) was found in those regions of the cell body located in the proximity of the active neurite junction. Interestingly, after the active phase ended (typically in less than 1 hour) the stiffness of these regions decreased to the initial values displayed before the neurite extension started. This phenomenon is illustrated in Fig 3. Fig 3 *a* shows a cell which undergoes neurite extension with an active growth cone region, monitored via optical microscopy. Over the course of 15 minutes (duration of the force map acquisition) the cell shows a minor change in growth cone location for the top neurite, and a substantial increase in neurite length for the bottom neurite (compare Fig. 3 *a* and Fig. 3 *b*). The force map (Fig. 3 *e*) and the histogram of elastic modulus (Fig 3 *g*: dash line) are compared with those measured on the same cell (Fig 3 *f* and *g*: solid line) during a subsequent phase of no growth. Fig. 3, *c* and *d* display optical images of this later phase, where the neurites exhibit no visibly active growth cone and no neurite length extension over the course of the force map. Overall, we see a significant stiffening of the area local to the neurite junction during neurite outgrowth (Fig. 3 *e*) as compared to the stiffness map during no extension (Fig. 3 *f* ). We also see a general shift in the histogram plot (Fig. 3 *g*) towards stiffer values during neurite extension, with an increase in highest value measured, as well as number of stiff points at or above 400 Pa. The average over all measured values of elastic moduli on the cell shifts from $(192 \pm 11)$ Pa during extension, down to $(142 \pm 6)$ Pa during the subsequent passive phase. This global value shift of over 30% is well above the typical $\leq 14\%$ variation in the average of the elastic modulus for a non-growing cell between two subsequent maps (See Fig. S3). Similar results were obtained for all the cells that exhibited active growth during force map acquisition (N=5 cells, see Fig. S4 in the Supporting Materials for an additional example).

## Effect of Taxol and Nocodazole on cell body elasticity

Taxol is a well-studied drug with known microtubule stabilizing effects in neurons (40, 41). By mapping the elasticity of live neuronal cell bodies before and after the addition of Taxol we were able to determine the effect of microtubule stabilization on live cell body elasticity. We performed elasticity maps on live cells at 37º C, and then exchanged the cell media with new media containing Taxol at a concentration of 10 μM. The cells were then incubated in the new media for a minimum of 20 minutes, and the new elasticity maps were performed on the same cells as before.

Fig. 4 *a* and *c* show respectively optical and force map images of a cell in the passive state (no neurite extension) before the addition of Taxol, while Fig. 4, *b* and *d* show the corresponding images of the same cell after a 90 minute incubation in media containing 10 μM Taxol. The cell undergoes both an overall increase in global stiffness, and a clear increase in stiffness local to the neurite junction. The histogram in Fig. 4 *e* shows an increase in both the highest stiffness measured, and number of points with elastic moduli above 400 Pa. The average value for the elastic modulus on the cell increases by more than 30% (from 229± 35 Pa, pre-Taxol, with no neurite extension, up to 304± 43 Pa after Taxol). A similar effect was observed on 3 additional cells, with an increase in average elasticity ranging from 33% to 180% (see Fig. S5 in the Supporting Materials).

We have also used the drug Nocodazole in an attempt to measurably disrupt the microtubules of the cell and the process of neuritogenisis. We used a similar procedure as that with Taxol but flushed the chamber instead with 10 nM nocodazole (Sigma-Aldrich, St. Louis, MO) in media. The results indicated that cells subjected to 10 nM nocodazole died with a substantially increased rate (6 out of 8 cells), and the surviving cells showed no marked decrease in cell body stiffness (See Fig. S6 in the Supporting Materials). This is consistent with findings in literature which indicate that while nocodazole disrupts neuritogenisis and increases cell mortality, it does not measurably decrease microtubule aggregations present in the cell soma (42).

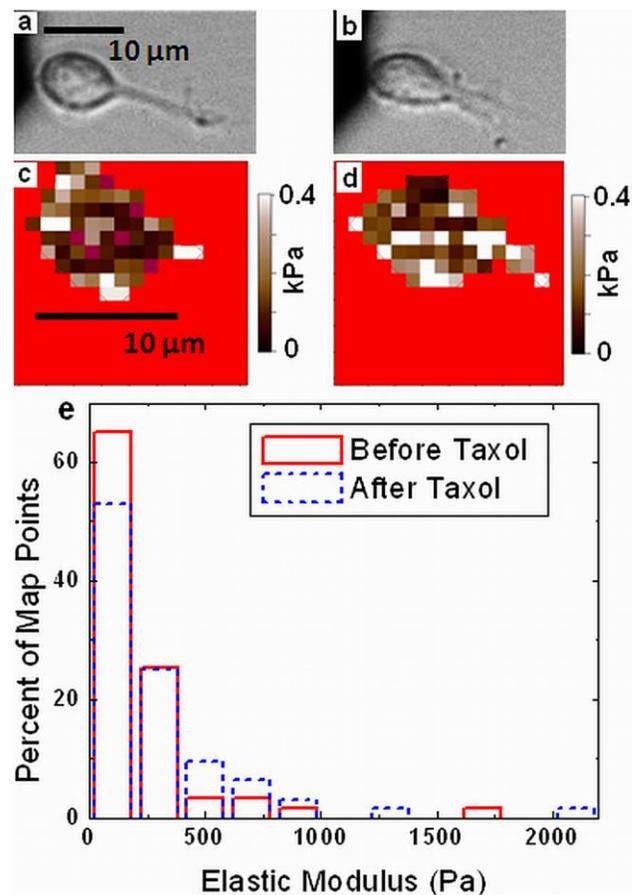

FIGURE 4 (a) Optical image of a live cortical neuron, which is not undergoing neurite extension. (b) Optical image of the same cell as in (a) shown 90 minutes after addition of 10 μM Taxol. Scale bar is the same for (a) and (b). (c), (d) Elasticity maps for cell shown in (a) and (b) respectively. (e) Histogram of percent of map points in each elastic modulus bin (see Materials and Methods) for the maps shown in (c) (solid line) and (d) (dashed line). Scale bar same for both maps. The average elastic modulus value increases by 33% after the addition of Taxol. Similar results seen on 3 additional cells (see Supporting Materials).

## Stiffening by neurite extension vs. Taxol addition

The addition of Taxol to live cortical neurons yields a similar pattern and magnitude of stiffening to that of active neurite extension. To further investigate this similarity, we compared the results of elasticity maps of untreated cells during active growth to those of the same cells after incubation with 10 μM Taxol. Fig. 5 *a* and *b* show active growth in the observed cell during map acquisition *before* the addition of Taxol. Two maps were taken on this cell during active extension (one of which is shown in Fig. 5 *e*), and another two maps were taken *after* the addition of Taxol (one of which is shown in Fig. 5 *f*). The data from the two elastic maps acquired during neurite extension were combined to produce the extension data set (Fig. 5 *g*, solid line), and the two Taxol maps were combined to produce the Taxol data set (Fig. 5 *g*, dashed line). We note that in both elasticity maps (Fig. 5, *e* and *f*) there is a similar distinctive high stiffness region local to the neurite extension, despite the clear lack of active extension during the Taxol maps (Fig. 5, *c* and *d*). We note further that the histograms show a similar maximum stiffness value, as well as a similar number of measured points at or above 400 Pa. Additionally, the average elastic modulus for the neurite extension maps is (255 ± 18) Pa, which falls nearly identical to the average elastic modulus for the Taxol maps of (247 ± 16) Pa.

## Identification of measured intracellular components by combined AFM and fluorescence microscopy

We have shown that active neurite extension in cortical neurons increases the measured elasticity in live neuronal cells near the active neurite junction, and further that this increase in elasticity is closely mirrored by the elasticity increase caused by the addition of the

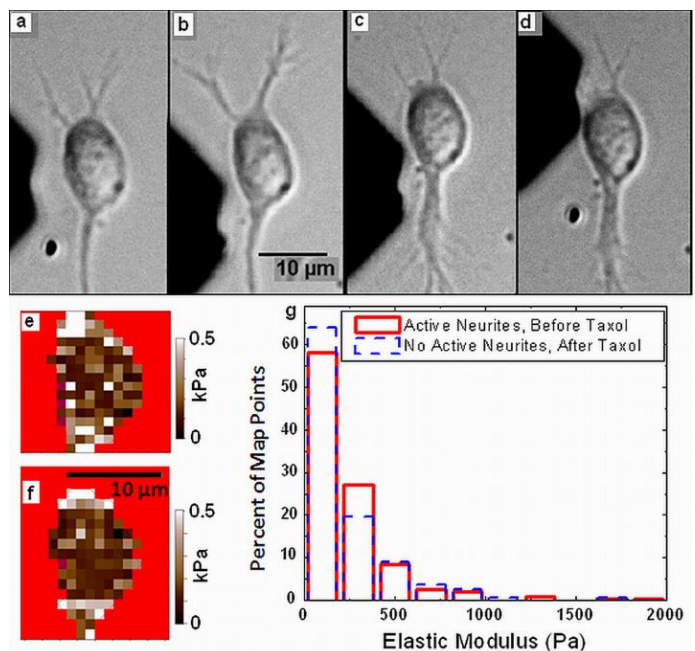

FIGURE 5 (a, b) Optical images of a live cortical neuron undergoing active neurite extension before the addition of Taxol; (a) shows the neuron before and (b) after the force measurements shown in (e); the upper neurite shows length extension during map acquisition (15 minutes). (c, d) optical images of the same cortical neuron shown in (a-b) taken *after* the addition of 10μM Taxol and incubation of 20 minutes. (c) Displays the neuron before and (d) after the measurements shown in (f). Scale bar shown in (b) is the same for all images (a-d). (e) Elasticity map for the case shown in (a-b); (f) Elasticity map for the case shown in (c-d). (g) Histogram of percent of total points in each elastic modulus bin (see Materials and Methods) for maps taken on the cell shown in (a-d). Solid line: data from 2 separate elastic maps acquired during neurite extension. Dashed line: data from 2 separate elastic maps acquired after addition of Taxol (both maps were measured >20 minutes after exposing the cell to Taxol). Average elastic modulus values between maps (e, f) differ by only 3%.

microtubule stabilizing drug Taxol. To further investigate these effects, we correlated AFM elasticity maps of living and fixed cortical neurons to fluorescence maps indicating regions of high microtubule or F-actin concentration. Fig.6 *a* shows a live cortical neuron stained for microtubule concentration. The image shows high microtubule concentration (i.e. high fluorescence intensity) along the top of the cell, as well as a significant aggregation local to the right-hand neurite junction. Fig. 6 *b* shows the AFM acquired elasticity map of this same cell immediately after the optical image acquisition. We see a direct matching between the regions of high elastic modulus (light areas) on the elasticity map, and the regions of high microtubule density observed through fluorescence. Similar correlations are seen for all cells stained for microtubules (N=6 cells, see Fig. 1 *a, b* and Fig. S7 in the Supporting Materials for additional examples).

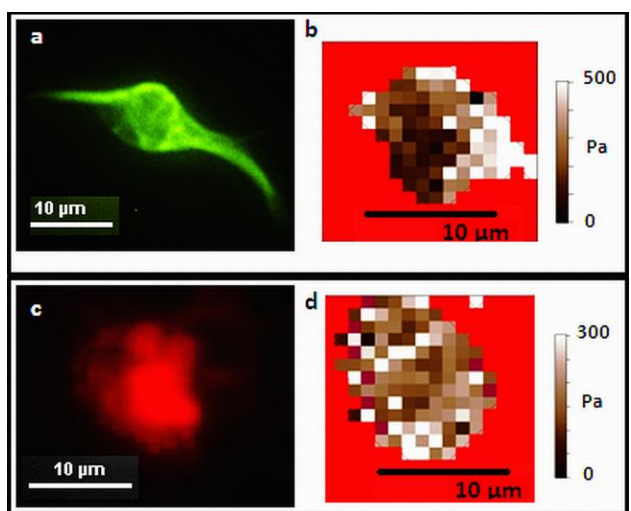

FIGURE 6 (a) FITC fluorescence image of live cortical cell stained for microtubules with 50nM Tubulin Tracker Green. (b) Elasticity map of cell shown in (a). The cell regions with high microtubule concentration (bright areas) in (a) correspond to the high stiffness regions shown in (b). Similar correlations were obtained for 5 additional cells (see Fig. 1 *a, b* and Fig S7 in the Supporting Materials). (c) Texas Red fluorescence image of cortical cell after being fixed and stained for F-actin with Alexa Fluor® 564 Phalloidin. (d) Elasticity map of cell shown in (c) prior to fixing. There is no correlation between the cell regions with high actin concentration (bright regions in (c)) and the cell regions that display high stiffness in (d). Similar results were obtained in 4 additional cells (see Fig S7 in the Supporting Materials).

Fig. 6 *c* shows a fixed cortical neuron stained for F-actin. We see in this image a bundle of higher density F-actin covering the majority of the lower right region of the cell. Fig. 6 *d* shows the elasticity map of this same cell still alive after elastic mapping and directly before (<20 minutes) fixing, showing, as in previous maps, regions of higher and lower elastic modulus. We note that the concentrated distribution of actin in the lower right of the body does not correspond to the regions of high stiffness in the elasticity map. Additionally, many high stiffness regions are seen in areas of the cell where F-actin aggregation is low. Similar results were obtained for all cells stained for actin (N=4 cells, see also Figure S7 in the Supporting Materials).

Finally, to determine if the increase in stiffness of the cell body during neurite extension corresponds to an increase in axonal tension, we acquired elasticity maps of cells *before and after active neurite growth* in media containing 10 μM Blebbistatin (see Fig. S8 in the Supporting Materials). Blebbistatin is a well-known inhibitor of nonmuscle myosin II, which was shown to dramatically reduce traction forces and axonal tension (16). In our experiments, all cells (N=3) that grow in the presence of Blebbistatin display an increase in stiffness between 30-55% during growth, with most stiffening regions occurring above 400 Pa. This increase is the same as the median change in stiffness values measured for cells which display active growth *in the absence* of Blebbistatin (see Fig. 3), which further indicates that microtubule aggregation (and not axonal tension) is primarily responsible for the observed stiffening during cell growth in our experiments. Our data also show that Blebbistatin does not significantly affect the cell stiffness in the passive (no-growth) state, or the aggregation of microtubules (See Fig. S9 in the Supporting Materials).

## DISCUSSION

We have shown the first direct comparison between elasticity maps on rat cortical, mouse P-19 derived, and chick DRG neurons. The overlap between elasticity values measured on P-19 derived and cortical neuron cell bodies (Fig. 1) yields additional evidence that P-19 derived neuronal cells are a good model system for cortical type neurons (38).

The difference in elasticity values of DRG neurons vs. cortical neurons (Fig. 1) is interesting in the context of their native environments. Cortical cells live in one of the softest tissue environments in the body, with bulk tissue modulus values reaching only a few hundred Pa (18, 30, 43). DRG neurons, in contrast, originate in nerve bundles along the spinal column, existing in and interacting with an environment both stiffer and more varied than the weak and relatively mechanically homogeneous cortex. The spinal column itself, for example, has shown bulk modulus values for different systems of the order of 10 kPa (43). The mechanical stiffness of a cell and its ability to generate forces are linked inexorably with the ability to manipulate or sense stiffness within that environment. A cell required to sense and function in a very weak and relatively homogeneous environment need not have the mechanical rigidity to sense and manipulate substantially stiffer surfaces. Koch, et al. have shown that the growth cones on hippocampal neurons generate extremely weak traction forces, and are insensitive to increases in substrate rigidity of 150 Pa and above (16). They have also shown that DRG neurons generate vastly larger traction forces than do hippocampal neurons, and are most sensitive to substrate stiffness in the range of 0.45-3 KPa (16).

We find it very relevant to this discussion that the DRG's, which generate larger traction forces and must sense and manipulate a stiff and varied environment, are stiffer than the weakly interacting and mechanically insensitive neurons from the brain. It is also important to note that the DRG's are particularly sensitive to substrate stiffness changes within the range of elasticity values reflected in their own elasticity maps. The maximum sensitivity range established by Koch and collaborators (16) of 0.45-3 KPa (with a particularly dramatic jump in preference between 450 and 1000 Pa) aligns well with our medium (0.5 KPa) and highest (2.9 KPa) values measured on the elasticity maps of DRG neuronal bodies (Fig. 1) as well as the typical soma average of 0.9 Pa.

We have additionally observed that all three types of neuronal cells show similar elasticity distributions (within one standard deviation) when plated on PDL, fibronectin, or laminin coated glass. This finding is of particular importance, as the role of matrix molecules on cell adhesion, propagation and differentiation is a robust area of study. Cell-matrix interactions are usually mediated by integrin-specific ligands that upregulate various pathways involved in cell responses to surfaces. Our data rules out large scale effects on the cell body elasticity for the three types of substrate studied here. However, since our data shows sizeable fluctuations in the measured elastic moduli due to cell to cell variation, we cannot exclude smaller scale effects of the substrate coating on the cell elasticity. We note that further studies of cell elasticity on various substrates could provide an important discriminator for comparing changes in cell propagation (e.g., via biochemical mediators) vs. changes in cell mechanics (e.g., demonstrated here via microtubule disruptors). Elucidating the roles of different inputs to nerve cell functions could also provide critical control points for future modes to direct the process in selective ways.

We also present the first use of AFM elasticity mapping to monitor differences in neuronal cell body elasticity over time, showing its power to resolve changes internal to live and healthy cells due to neurite extension and drug response. Additionally we use the unique power of combined AFM and fluorescence microscopy to analyze the internal cellular components responsible for these changes. Specifically, we have identified the areas of high elastic modulus measured in the cell bodies of cortical neurons as areas of high microtubule density rather than concentrated regions of F-actin (Fig 1, Fig. 6, Fig. S7). We have shown that the addition of Taxol to a live cell increases the stiffness in these areas to a degree easily measureable by AFM elasticity mapping (Fig. 4, Fig. 5, Fig. S5) further identifying and confirming the active effect of Taxol on microtubules in live neurons. We have additionally discovered a dynamic and reversible stiffening of the cell body local to neurite junctions in response to active neurite extension (Fig. 3, Fig. S4). This stiffening effect has been shown to be of comparable magnitude to changes induced by the addition of Taxol (Fig. 5), and by our fluorescence results (Fig 1, Fig. 6, Fig. S7) we can identify these significant increase in stiffness as due to microtubule dynamics rather than changes in F-actin concentration. In addition, cell treatment with Nocodazole (a drug known to disrupts neuritogenisis, but which does not measurably decrease microtubule aggregations in the cell soma) and Blebbistatin (known to dramatically reduce tension forces generated during axonal elongation) show no significant effect on the elasticity maps of the cells. Also, Blebbistatin does not reduce the stiffening effect observed during growth, further supporting our conclusion that microtubules are primarily responsible for the observed increase in stiffness in our experiments. These results are supported by current axonal growth models, which demonstrate that microtubules have major roles in the process of axonal extension. These models show that although actin filaments are remodeled very fast in response to guidance cues, axons cannot move forward without the steering and mechanical stabilization induced by microtubules (1, 44). In particular we associate the local increase in stiffness near the area of neurite extension to the formation of quasi-stable bundles of microtubules which enter the axonal shaft. The decrease in stiffness observed after the extension phase suggest a de-polymerization of these microtubule structures, at least in the case when neurons does not form functional connections with nearby cells. Additional biochemical and mechanical studies, especially on cells that form functional connections would help to expand our understanding of the either distinctive or synergistic roles of these various cytoskeletal inputs to mechanics as studied here.

Our findings also suggest new strategies to consider with regard to directing nerve cell growth in 2D and 3D systems. For example, gradient biomaterials where mechanics, ECM factors and cytoskeletal disruption factors, are appropriately positioned may provide improved directionality of nerve cells. This directionality could go beyond that currently achieved through surface patterning or macroscale gradients with nerve growth factors. Monitoring the mechanics of the cells in response to such treatments may also provide real-time information regarding nerve cell responses to selective chemicals and toxicants.

## ACKNOWLEDGEMENTS

The authors thank Amy Hopkins and Dr. Min Tang-Schomer for their assistance and technical guidance on the methods. We thank Gregory Frost and Dr. Steve Moss's laboratory at Tufts Center of Neuroscience for providing embryonic rat brain tissues. We also thank Prof. Leon Gunther and Prof. Tim Atherton (Tufts Physics) for useful discussions.
This work was supported by the National Science Foundation (grant CBET 1067093), by the Tufts Collaborates Seed Grant, and the National Institute of Health (grants AR005593 and EB002520).

**Supporting Material for:**

**Elasticity Maps of Living Neurons Measured by Combined Fluorescence and Atomic Force Microscopy**

Elise Spedden*†, James D. White*†‡, Elena N. Naumova §, David L. Kaplan‡, Cristian Staii*†
* Department of Physics and Astronomy, Tufts University, †Center for Nanoscopic Physics, Tufts University, 4 Colby St, Medford, MA, USA; ‡Department of Biomedical Engineering, and Department of Chemical Engineering, Tufts University 4 Colby St, Medford, MA, USA, § Department of Civil and Environmental Engineering, Tufts University, 200 College Avenue, Medford, MA, USA

**Text S1:**

Dorsal root ganglia were surgically isolated from day 9 chick embryos and placed in Hanks Balanced Salt Solution (Life Technologies). The ganglia were incubated in 0.25% Trypsin (Life Technologies) for 30 minutes at 37°C, centrifuged, and the cell pellet was re-suspended in Dulbecco's Minimum Essential Media (Life Technologies) (high glucose supplemented with GlutaMAX (Life Technologies), penicillin/streptomycin (pen/strep) (Life Technologies) 1%, and fetal bovine serum (FBS) (Life Technologies) (2%)). The cells were mechanically dissociated and the suspended cells were added to a cell culture dish, and incubated for 30 minutes allowing adsorption of astrocytes to the dish surface. The remaining media containing neurons was removed. The cells were counted, plated at 200,000 cells per 3.5 cm culture disk, and grown for 4 days.

Rat cortices were obtained from embryonic day 18 rats (Tufts Medical School). The corticies were incubated in 5 mL of trypsin at 37ºC for 20 minutes, then the trypsin was inhibited with 10 mL of neurobasal medium (Life Technologies) supplemented with GlutaMAX, b27 (Life Technologies), and pen/strep, containing 10 mg of soybean trypsin inhibitor (Life Technologies). The neurons were then mechanically dissociated, centrifuged, the supernatant removed, and the cells were resuspended in 20 mL of neurobasal medium containing L-glutamate (Sigma-Aldrich, St. Louis, MO). The neurobasal media was implemented to support neuronal growth without the use of serum, thereby reducing glial cell proliferation. The cells were re-dispersed with a pipette, counted, and plated at a density of 250,000 cells per 3.5 cm culture disk.

P-19 mouse teratocarcinoma stem cells (American Type Culture Collection, Manassas, VA) were cultured using αMinimum Essential Media (Life Technologies, Grand Island, NY) supplemented with FBS (2%), calf bovine serum (CBS) (7.5%) (Life Technologies), and pen/strep (1%). Differentiation was accomplished by incubating the P-19 cells in ultra-low-adhesion cell culture flasks in the presence of retinoic acid (Sigma-Aldrich, St Louis, MO) (2 μM). After 2 days, cell clumps were mechanically dissociated and re-suspended in fresh medium, also containing 2 μM retinoic acid. On day 5, the cell clusters were mechanically dissociated, and plated in fresh medium at a density of 50,000 cells per 3.5 cm culture disk. After an additional 24 hours, cytosine arabinoside (Sigma-Aldrich) was added to remove any undifferentiated cells. The cells were incubated for a minimum of 3 days.

**Text S2:**

Before measurement on a new sample, each cantilever was calibrated both in air over cleaned glass (to determine accurate lever sensitivity and spring constant values), and in the sample medium (to determine the new lever sensitivity value in fluid.

The in-fluid calibration was performed by two methods to verify accuracy: a thermal calibration (built in to the MFP3D software) that uses the previously determined spring constant value to extract the new sensitivity value, and a deflection vs. z-movement curve calibration on a clean glass area of the sample free from cells or debris. Generally, the two calibration methods gave consistent results and the calibration was considered accurate.

To verify that all cells to be measured were alive, 10 minutes time lapse videos were taken of each cell set before AFM measurements. A 40x objective was used to optically locate the AFM cantilever above each cell and 16 X 16 μm maps of individual force vs. indentation curves were taken on each cell with a resolution of 1μm between points. To limit energy dissipation due to viscoelastic effects the cantilever z velocity was kept at 2 μm/sec, with a maximum cantilever deflection between 5-10nm. A number of 5 to 7 well-adhered cells were mapped for each experimental condition (three cell types:

chick DRG, mouse P-19 derived, and rat cortical neurons, three surface coatings for each cell type: PDL, laminin, and fibronectin). To assess adherence, during force map acquisition each cell was monitored visually via the 40x objective to rule out any cells that underwent lateral slipping under the force from the cantilever. Cell edges were determined using height data, with all points not on the cell body excluded.

For force data analysis, highly noisy and poorly fitting curves (generally less than 10% for each force map) were excluded from the data. Based on height information all data on areas outside of the cell body region was also excluded. To verify the implemented MFP3D analysis, several curves on multiple samples were also fitted independently by the authors using the Hertz model equation. In this case, the elastic modulus of a force vs. indentation curve was extracted using Sneddon's modification of the Hertz contact model for a $30^0$ conical indenter:

$$F = \frac{E}{(1-\nu^2)} \frac{2\tan(\alpha)}{\pi} \delta^2 \quad (1)$$

where F is the force, E is the elastic modulus, α is the half-angle of the conical indenter, and δ is the indentation depth. For the Poisson ratio (ν) we use 0.33. For curves that were fitted by both methods (Built-in MFP3D software and by-hand analysis in Origin) the results typically agreed to within 90%.

**Table S1:** p values for the 1-way ANOVA tests for the effect of surface coating; PDL= Poly-D-lysine, LN= Laminin; FN= Fibronectin. The Top 10%, Middle 30% and Bottom 10% values are defined in the main text (see Materials and Methods). The large p values for the majority of combinations of neuron types, surface coatings and ranges of values for elastic modulus show that the cell stiffness is not significantly affected by the surface coating (see also Figure S2).

|          | High PDL v LN | High PDL v FN | High LN v FN | Med PDL v LN | Med PDL v FN | Med LN v FN | Low PDL v LN | Low PDL v FN | Low LN v FN |
|----------|---------------|---------------|--------------|--------------|--------------|-------------|--------------|--------------|-------------|
| P19      | 0.79418       | 0.79312       | 0.95678      | 0.74986      | 0.87033      | 0.7965      | 0.91614      | 0.9067       | 0.73849     |
| DRG      | 0.14406       | 0.72038       | 0.37745      | 0.90364      | 0.04925      | 0.13952     | 0.83024      | 0.3037       | 0.32164     |
| Cortical | 0.18441       | 0.17243       | 0.70251      | 0.93726      | 0.07685      | 0.17415     | 0.3498       | 0.29583      | 0.05512     |

**Figure S1:**

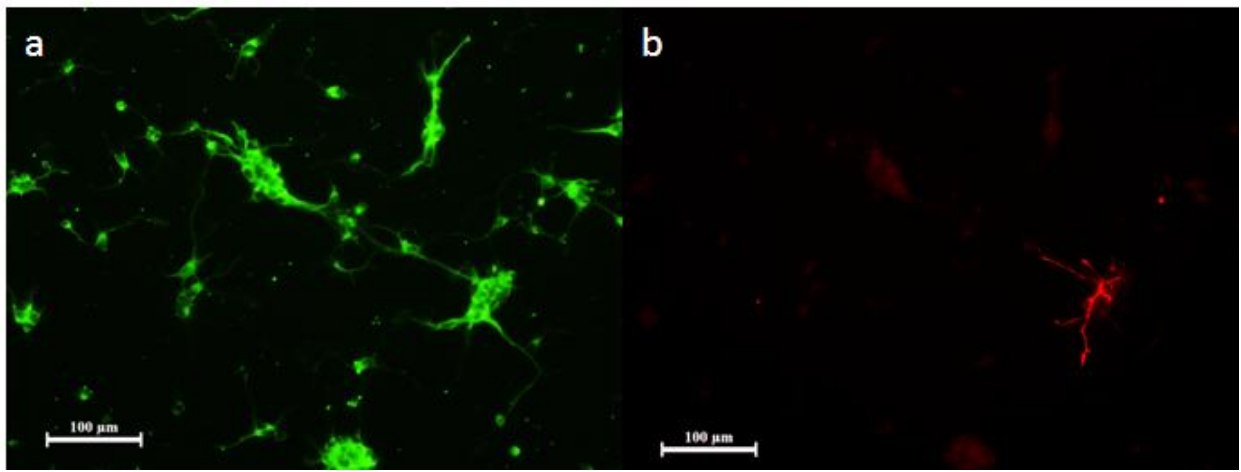

FIGURE S1 Immunostaining experiments on the same region of cortical neuronal cell culture using (a) Anti-β-tubulin III (Sigma Aldrich, St. Louis, MO) diluted 1:500 (neuronal marker) and (b) Anti-Glial Fibrillary Acidic Protein (Sigma Aldrich, St. Louis, MO) (glial cell marker) antibodies. The image indicates cultures of high neuron cell purity. For all measured samples, cortical neurons were further identified via typical morphology. In addition, for all measured DRG neurons we have measured only those cells that display very long processes (≥100μm), which are representative for DRG neurons. P-19 derived neurons were chosen based on morphological similarity to cortical neurons and long processes that do not typically branch. All force maps on P-19 and DRG neurons (Fig 1 and 2 in the main text) were performed on this type of cells, for which all the processes were fully grown (no active growth state was observed on well-developed P-19 and DRGs).

**Figure S2:**

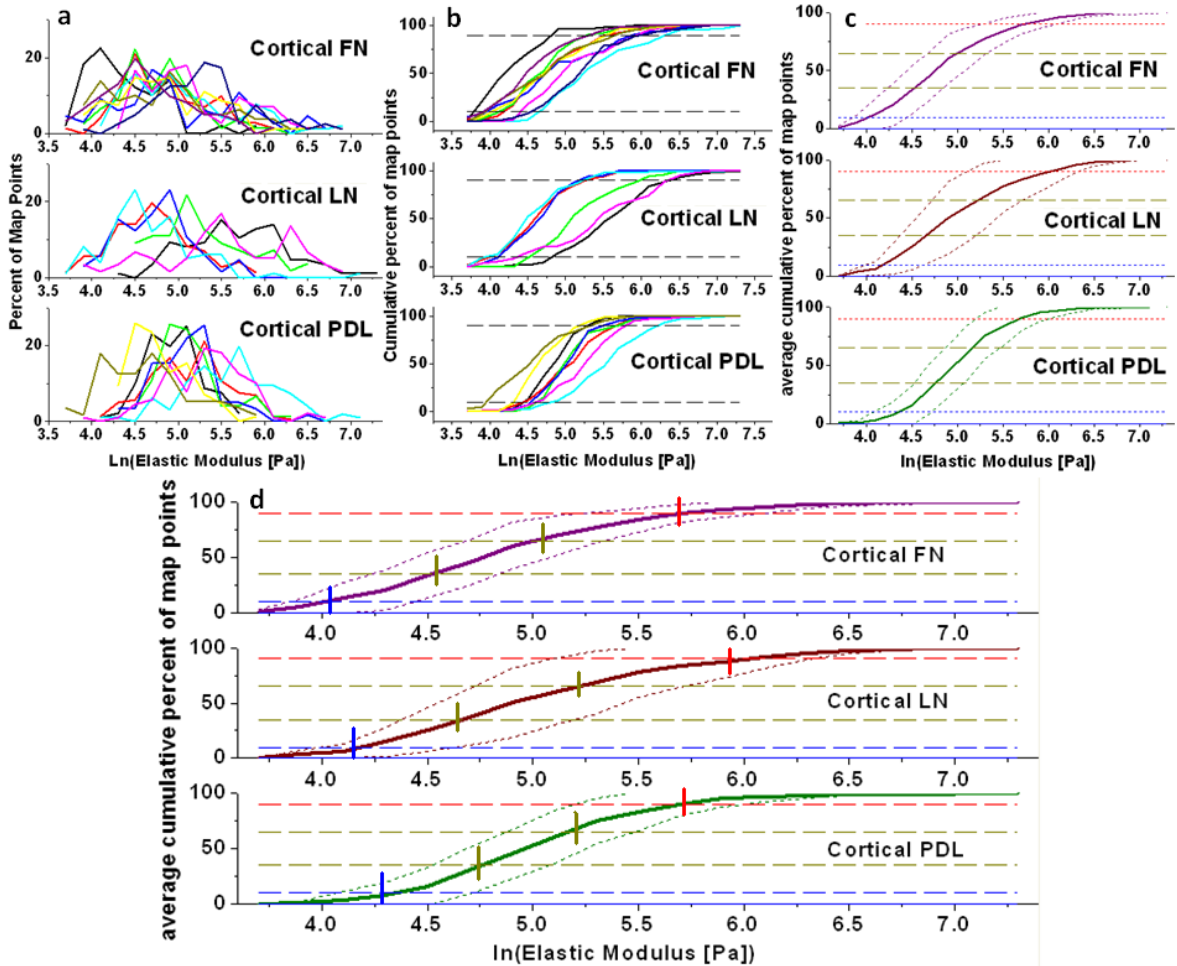

FIGURE S2 (a) Complete individual distributions of map points for the elastic modulus, compiled for each cell, within each surface coating; *top*: fibronectin (FN), *middle*: laminin (LN), *bottom*: poly-D-lysine (PDL). Distributions are displayed as percent of map points within each "bin" plotted vs. the natural log of the measured elastic modulus (Pa) for each bin. To best represent each data set as a distribution, optimal Bayesian binning was applied to log-transformed data. Individual lines represent the distributions from different cells. (b) Cumulative distributions for elastic modulus (i.e. sums of the data points from (a) up to a given bin) vs. the natural log of the measured elastic modulus (Pa) for each bin. Dotted black lines show the cutoff limits for $10^{th}$ and $90^{th}$ percentiles, or the top 10% of data and bottom 10% of data, respectively. (c) Average cumulative distributions shown for each surface as the solid line. Dashed yellow lines show the cutoff limits for 35th and 65th percentiles, or the middle 30% of data. Dotted lines above and below each cumulative distribution illustrate +/- 1 standard deviation (SD) that form a confidence area for each data set. *The confidence areas of cumulative distributions for each surface type overlap along the whole range of values, indicating a very low probability for a surface-dependent effect outside of 1 SD from the average distribution.* (d) Expanded version of the plot shown in (c) with the vertical bold markings indicating the points at which the cumulative average curves cross into or out of the bottom 10% region (blue dashes), the middle 30% region (yellow dashes), or the top 10% region (red dashes). The good level of vertical alignment between these cross points indicates that it is appropriate to use the top 10%, middle 30% and lowest 10% areas as representative of those regions of the data. Similar results are obtained for cumulative distributions calculated for P-19 and DRG cells cultured respectively on FN, LN and PDL coated glass surfaces.

**Figure S3:**

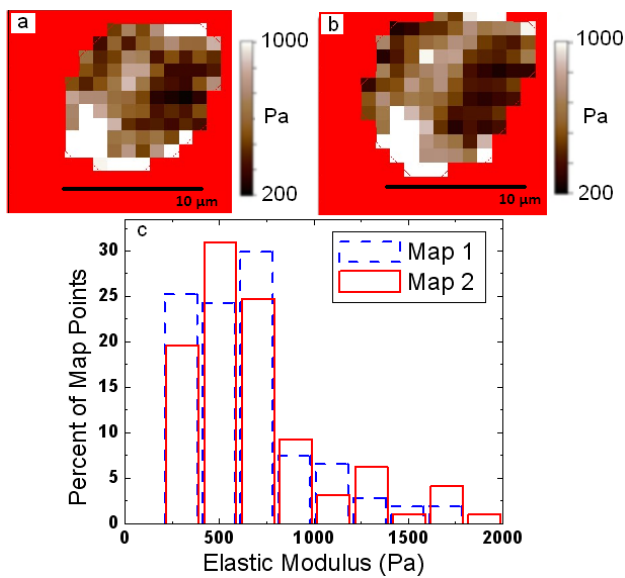

FIGURE S3 (a) Elasticity map of unmodified live cortical cell not undergoing active neurite extension. Histogram of percent of map points in each elastic modulus bin shown in (c) (blue dash). (b) Subsequent elasticity map of the same cell as shown in (a), in the same conditions (no active neurite extension, no chemical modification) taken after 45 min. Histogram of percent of map points in each elastic modulus bin shown in (c) (red solid). Average elastic modulus values between maps (a, b) differ by only 6.5%. Similar results were obtained in 5 other live unmodified cells with no active neurite extension, and with maps taken between 20 minutes and 2 hours apart. The differences in average elastic modulus values obtained from these maps range between a minimum of 3% and a maximum of 14%.

**Figure S4:**

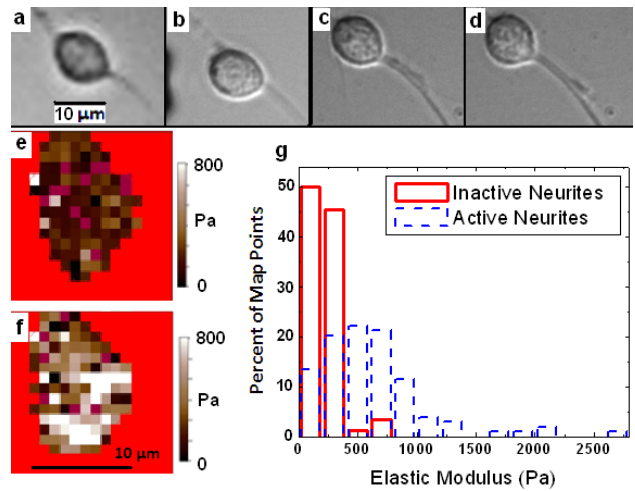

FIGURE S4 (a, b) Optical images before (a) and after (b) force measurements of a live cortical neuron *not undergoing* active neurite extension during 15 minute force-map (passive phase). (c, d) Optical images before (c) and after (d) force measurements of same live cortical neuron at a later time actively undergoing neurite extension (active phase), seen as an increase in the length of the newly visible neurite in the lower right. Scale bar shown in (a) is the same for all images (a-d). (e) Elasticity map for the passive phase shown in (a-b). (f) Elasticity map for the active extension phase shown in (c-d). Scale bar shown in (e) is the same for both maps. g) Histogram of percent of total map points in each elastic modulus bin (see Materials and Methods). Dashed line: data for active extension state. Solid line: data for the passive state. The average elastic modulus value increases by 175% during growth. For all cases where neurons display active neurite extension, we *always* measure an increase in stiffness in those regions of the cells located in the proximity of the active neurite. The combined data from these regions for all (N=5) cells, accounts in average for more than 75% of the overall increase in the stiffness of the cell body observed during growth (and could be up 90% of the overall increase for some individual cells, as shown in Fig S4 *e, f*). The data for all cells (N=5) shows that the stiffening of the cell regions close to active neurites is the primary effect that accounts for the overall increase in cell stiffness. However, we also find other regions of the cell that stiffen during the growth phase (shown for example in Fig. 3 *e*). The contribution of these regions to the overall increase in stiffness is typically less than 25%.

**Figure S5:**

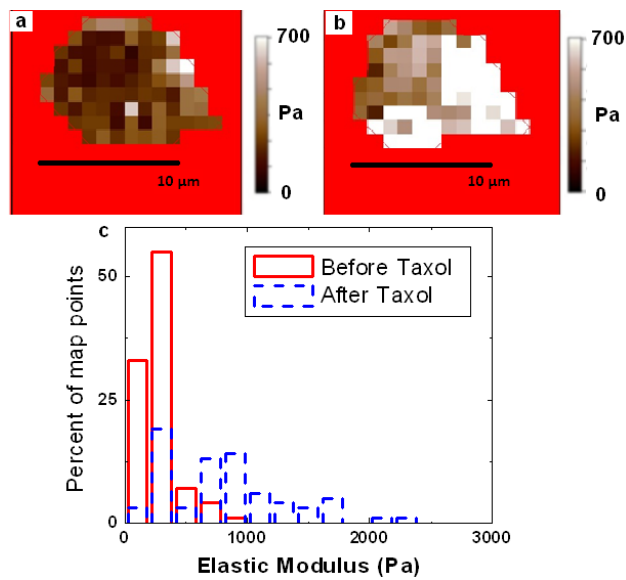

FIGURE S5 (a) Elasticity map of a live cortical neuron, which is not undergoing neurite extension. (b) Elasticity map of the same cell as in (a) shown 90 minutes after addition of 10 µM Taxol. Scale bar is the same for (a) and (b). (c) Histogram of percent of map points in each elastic modulus bin (see Materials and Methods) for the maps shown in (a) (solid line) and (b) (dashed line). Scale bar same for both maps. The average elastic modulus value increases by 180% after the addition of Taxol. Similar results seen on 3 additional cells.

**Figure S6:**

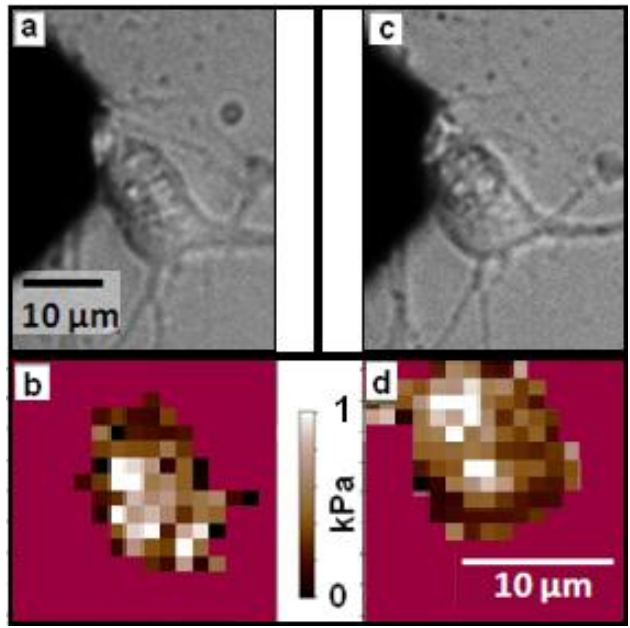

FIGURE S6 (a) Optical image of live cortical cell. (b) Elasticity map of cell shown in (a). (c) Optical image of same cell shown in (a) after application of 10 nM Nocodazole (Sigma-Aldrich, St. Louis, MO). (d) Elasticity map of cell post-Nocodazole showing no appreciable change in overall cell stiffness. Similar results were seen on 1 other treated cell. Additional treated cells (6 out of 8) died before the acquisition of a second force map.

**Figure S7:**

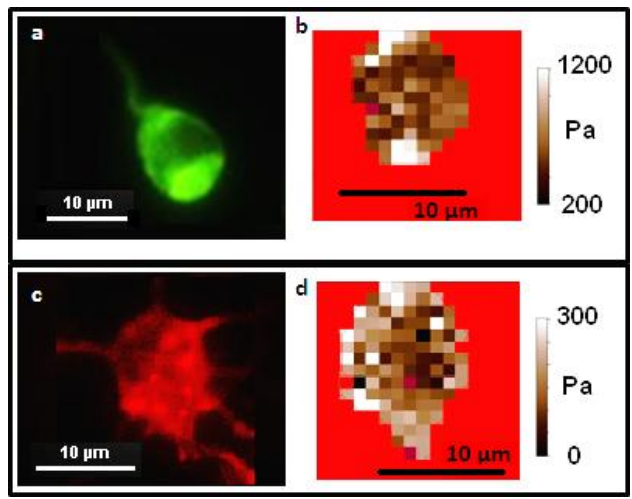

FIGURE S7 (a) FITC fluorescence image of live cortical cell stained for microtubules with 50nM Tubulin Tracker Green. (b) Elasticity map of cell shown in (a). The cell regions with high microtubule concentration (bright areas) in (a) correspond to the high stiffness regions shown in (b). Similar correlations were obtained for 5 additional cells. (c) Texas Red fluorescence image of cortical cell after being fixed and stained for F-actin with Alexa Fluor® 564 Phalloidin. (d) Elasticity map of cell shown in (c) prior to fixing. There is no correlation between the cell regions with high actin concentration (bright regions in (c)) and the cell regions that display high stiffness in (d).

**Figure S8:**

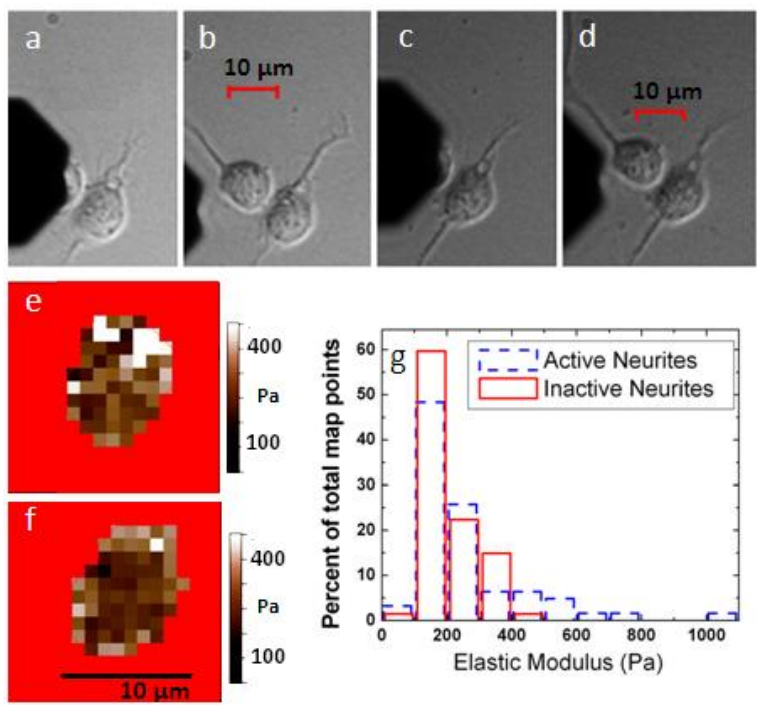

FIGURE S8 (a, b) Optical images before (a) and after (b) force measurements of a live cortical neuron undergoing active neurite extension during 15 minute force-map (*active phase*: seen as an increase in length of the extending top neurite). All measurements were performed in media containing 10 μM Blebbistatin) (c, d) Optical images before (c) and after (d) force measurements of same live cortical neuron (in media containing 10 μM Blebbistatin) at a later time *not undergoing* neurite extension (*passive phase*). Scale bar same for (a), (b), and same for (c), (d). (e) Elasticity map for the active extension phase shown in (a-b). (f) Elasticity map for the passive phase shown in (c-d). Scale bar shown in (f) is the same for both maps. g) Histogram of percent of total map points in each elastic modulus bin (see Materials and Methods). Dashed line: data for active extension state. Solid line: data for the passive state. Similar results were obtained for 2 additional cells. The results show a measured 30-55% increase in stiffness due to growth, which is a similar change to that seen in the majority of growing samples without Blebbistatin.

**Figure S9:**

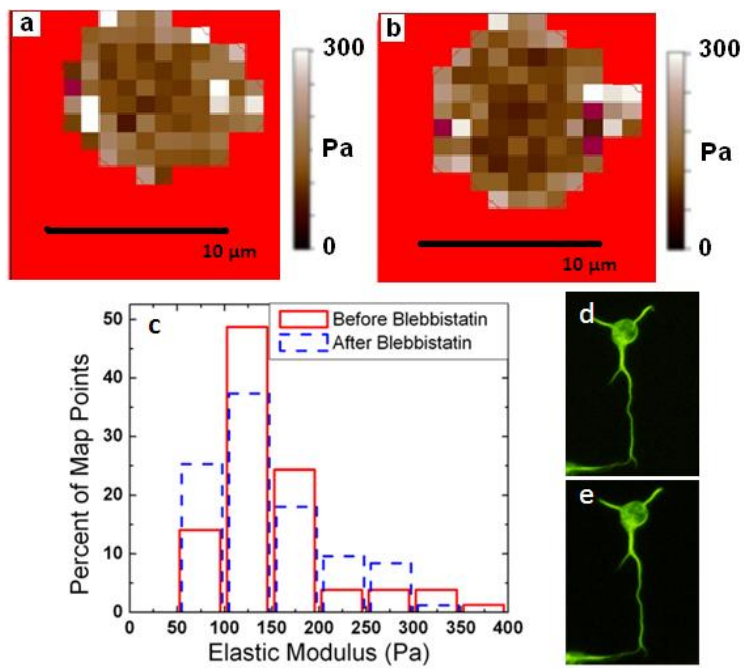

FIGURE S9 (a) Elasticity map of a live cortical neuron. (b) Elasticity map of the same cell as in (a) shown after application of 10 μM Blebbistatin. Scale bar is the same for (a) and (b). (c) Histogram of percent of map points in each elastic modulus bin (see Materials and Methods) for the maps shown in (a) (solid line) and (b) (dashed line). Scale bar same for both maps. Average elastic modulus values between maps (a, b) differ by only 3% indicating no baseline change in stiffness due to application of Blebbistatin. Similar results obtained from 2 additional cells. (d) FITC fluorescence image of live cortical cell (different cell from (a), (b)) stained for microtubules with 50nM Tubulin Tracker Green. (e) Fluorescence image of same cell as in (d) 30 minutes after application of 10 μM Blebbistatin indicating no change in tubulin aggregation after application of Blebbistatin.